\documentclass[manuscript]{copernicus}

\nolinenumbers
\usepackage{color}
\usepackage[T1]{fontenc}

\begin{document}

\title{Possible increased critical temperature T$_\mathrm{c}$ in anisotropic bosonic  gases
}

{\author[1,3]{R. A. Treumann}
\author[2]{Wolfgang Baumjohann$^a$}
\affil[1]{International Space Science Institute, Bern, Switzerland}
\affil[2]{Space Research Institute, Austrian Academy of Sciences, Graz, Austria}
\affil[3]{Geophysics Department, Ludwig-Maximilians-University Munich, Germany\protect\\
Correspondence to: Wolfgang.Baumjohann@oeaw.ac.at}

}

\runningtitle{Condensate formation}

\runningauthor{R. A. Treumann \& Wolfgang Baumjohann}

\received{ }
\pubdiscuss{ } 
\revised{ }
\accepted{ }
\published{ }


\firstpage{1}

\maketitle

  

\noindent\textbf{Abstract}.-- 
{A finite thermal anisotropy, if maintained for times longer than thermal relaxation times, may have a positive effect on the critical temperature in Bose-Einstein condensation of a dilute boson gas not in thermal equilibrium or quasi-particle fermi fluid consisting of spin-compensated electron pairs. It raises the transition temperature while increasing the condensate density. 
    } 

\section*{Introduction}

{There is a long-term interest in the superfluid {properties  \citep{kapitza1938,allen1938,landau1938,landau1949,ginzburg1958,pitaevskii1961,gross1961,schmitt2014} of massive boson gases when undergoing  Bose-Einstein condensation  (BEC). Critical transition temperatures for onset of BEC are close to absolute zero which complicates their investigation though advanced techniques   progressed in the past decades,  \citep{pethick2008} producing substantial insight into the dynamics of the condensate. Nevertheless, raising the transition temperature by some means would be advantageous. Similarly, in the  Bardeen-Cooper-Schrieffer (BCS) theory of superconductivity  \citep{BCS1957,BCS1957a,schrieffer1999,schrieffer1964} the quasi-particle Fermi fluid of electron pairs  \citep{cooper1956} below the superconducting transition temperature develops properties similar to a boson fluid. Increasing the transition temperature in this case is highly desirable for obvious practical reasons.}  The discovery of high-temperature  \citep{bednorz1986} superconductivity in suitably tailored metallic alloys opened up such a possibility, hopefully reaching into the domain close to room temperature  \citep{leggett2006}. Most recently experimental efforts progressed  \citep{kordyuk2012,kordyuk2018,drozdov2015}  by applying very high pressures of the order of $\sim200$ GPa to LaH$_{xx}$, \citep{drozdov2018,drozdov2015,goncharov2017,bianco2018}. Under these conditions the alloy becomes a promising BCS superconductor, while other theories of high-temperature superconductivity (HTSC)  \citep{somayazulu2019,dzhumanov1994,dzhumanov1994a,sachdev2016} usually deviate from BCS theory. 
 {An interesting theory of this type includes polaron interactions of fermions and afterwards between the boson-like properties of Cooper pairs  \citep{dzhumanov1995,dzhumanov1997}}. It has also been shown that high-temperature superconductivity in bismuthates {presumably is of the BCS form  \citep{wen2018} with an important modification caused by long-range Coulomb interaction effects  \citep{rice1981,meregalli1998a,meregalli1998} on the electron-phonon coupling  \citep{nourafkan2012,yin2013}. It leads to band expansion and the observed increase in the basic isotropic critical temperature $T_c$.} Still, the question remains, {whether in a suitably modified BCS theory the already achieved critical temperature  could, by some factor, be raised.} In the present Letter we suggest a theoretical possibility for such an increase of $T_c$ in the bosonic fluid which, to our knowledge, has so far not been considered yet. We refer to thermal non-equilibrium bosonic superfluid properties.} 

We assume the presence of a temperature anisotropy $\tau=T_\|/T_\perp\neq 1$, where $\|$ and $\perp$ indicate two orthogonal spatial directions. {Thermodynamically, such systems represent non-equilibria. In closed systems they would, after some finite time, thermalise and settle in an isotropic temperature state.  \cite{pethick2008} consider the case of producing an anisotropy in a boson gas trapped in a confining potential. In this case the system is closed. In a dense trapped gas the collisional thermalisation times at finite temperature are indeed short. However, by imposing a common chemical potential they assume that the gas is in thermal equilibrium, a condition we shall relax below. In order for an anisotropy to be maintained for a sufficiently long time, the system must be in thermal non-equilibrium and possibly open requiring an external agent which keeps it in that state. Recently \cite{orioli2019} numerically investigated another kind of non-equilibria near criticality in a diluted boson gas. One could consider, for instance, two sufficiently dilute counter streaming atomic flows or beams of same low temperature. In the dilute collisionless counter streaming case the relaxation times should be sufficiently long until reconstruction of a common thermal equilibrium will take place, in particular when the system is open. Under these assumptions,  the effective common temperature in the direction of the flows will be higher than perpendicular faking a temperature anisotropy. For the sake of simplicity and to investigate the effect of a temperature anisotropy, we treat the simpler case of a given temperature anisotropy here.} 

What concerns superconducting devices, so they are naturally in thermodynamic non-equilibrium states \citep{BCS1957,ginzburg1950a,ginzburg1986,bergeret2018} {where they may exhibit various kinds of inhomogeneities and anisotropies being placed under intense scrutiny. \citep{bergeret2018}}  Anisotropic-temperature momentum-space distributions in high temperature systems are well known in classical \citep{davidson1984,baumjohann1996} as well as quantum plasmas, \citep{wigner1932,manfredi2005,manfredi2015} {They are not in thermal equilibrium, giving rise to instability, momentum and energy transport and affecting the dynamics and stability of the systems} \citep{davidson1984,treumann1997,haas2000,manfredi2001,manfredi2018}.

{Though for the superconducting superfluid a similar analysis would doubtlessly be of interest,  we do not investigate any such effects on the behaviour of the superconducting quasiparticles.  Instead we assume that the state of the superfluid is for sufficiently long time subject to an anisotropic (and thus non-equilibrium) bose distribution, aiming on the effect the temperature anisotropy $\tau<1$ may exert on the (already achieved) critical temperature $T_c$ of the superfluid in BEC. In the superconducting case one should instead attempt the construction of  an anisotropic BCS distribution, contemplating in which way the anisotropy could be technically achieved. One may speculate whether, for instance, imposing a magnetic bottle configuration onto the lattice to trap the electrons would be technically feasible.} 

\section*{Anisotropic bosons}
The bosonic temperature anisotropy is arbitrarily assumed in the third direction of a local cartesian coordinate system $\{1,2,3\}$ with $T_3\equiv T_\|<T_\perp$ the parallel temperature, and $T_\perp$ the isotropic reference temperature in the two perpendicular directions. Defining $A_\tau=T_\perp/T_\|-1=\big(1-\tau\big)/\tau$, the distribution of states $n_\mathbf{p}$ with respect to the boson momentum $\mathbf{p}$  becomes
\begin{equation} \label{eq-1}
n_\mathbf{p}(T_\perp,A_\tau)-n_0=\Big(1+A_\tau\cos^2\theta\Big)\bigg\{\exp\frac{1}{T_\perp}\Big[\big(\epsilon_\mathbf{p}-\mu\big)\big(1+A_\tau\cos^2\theta\big)\Big]-1\bigg\}^{-1}
\end{equation}
where $\epsilon_\mathrm{p}=p^2/2m$,  $\tan\theta=p_\perp/p_\|$, and the zero momentum (ground)  state occupation $n_0=n_{\mathrm{p}=0}$ has been separated out.  (A proof is given in the Methods section.)
{This ground state still depends on the chemical potential $\mu$ and, through $A_\tau$ on the anisotropy $\tau$. Maintaining a thermal non-equilibrium state, as required, it is clear that the chemical potential in the two orthogonal directions must differ. Hence, the angle between the parallel and perpendicular chemical potentials is the same as that for the momenta  $\theta=\tan^{-1}\big(\mu_\perp/\mu_\|\big)$. } It is also clear that, like for all ideal boson gases, independent of the presence of anisotropy, the chemical potential $\mu\lesssim 0$ is negative with modulus approaching $\big|\mu\big|\to0$ at some low critical temperature $T_{\perp c}\big(\mu=0\big)$ which for massless bosons vanishes but is finite for massive bosons. This anisotropy-caused angular dependence complicates the problem. One may note that for $x=0$ the ordinary Bose distribution is recovered, independent of any anisotropy. Hence for all anisotropies the ground state occupation $n_0$ corresponds to the perpendicular direction at zero energy, thereby in retrospect justifying the choice of $T_\perp$ as reference temperature.

Putting $\mu=0$ and as usually integrating over energy  \citep{kittel1980,huang1987,lifshitz1998} yields an estimate for the critical temperature under the assumption of constant total particle number $\mathcal{N}$, constant volume $\mathcal{V}$, and density $N\equiv\mathcal{N}/\mathcal{V}$ from
\begin{equation}\label{eq-dens0}
N-N_0=\frac{2g}{\sqrt{\pi}\lambda_\perp^3}\int_0^1\frac{dx}{\big(1+A_\tau x^2\big)^\frac{1}{2}} \int_0^\infty\frac{y^\frac{1}{2}dy}{e^y-1}
\end{equation}
with $\lambda_{T_\perp}^2=2\pi\hbar^2/mT_\perp$ the (perpendicular) squared thermal wavelength, $y$ the dimensionless energy variable, and $x=\cos\theta$. The average  density $N_0\equiv N_{\epsilon=0}$ of particles in the zero energy ground state $n_0$ refers to the Bose-Einstein condensate (BEC). No distinction has been made  in the degeneracies $g$ of the parallel and perpendicular energy levels. 

Performing the $y$-integration yields the well-known isotropic result $\zeta\Big({\textstyle\frac{3}{2}}\Big)\Gamma\Big({\textstyle\frac{3}{2}}\Big)$ which is just a number for $\mu=0$. The integration with respect to $x$ has decoupled and we have the result
\begin{equation}
N-N_0=\frac{g}{\lambda_{T_\perp}^3}\zeta\Big({\textstyle\frac{3}{2}}\Big)\frac{\mathrm{sinh^{-1}}\sqrt{A_\tau}}{\sqrt{A_\tau}}
\end{equation}
This behaves regularly for $\tau=1$ or $A_\tau=0$, as can be shown applying l'Hospital's rule, which yields the isotropic classical result. (We will generalise this to finite values of the chemical potential in the Methods section.) On the other hand, for $A_\tau\gg1$ which implies that $\tau\ll1$ the right-hand side becomes very small thus indicating that for large anisotropies $N_0(\tau\gg1)\to N$ such that the fraction of particles which condensate in the degenerate ground state increases.

\section*{Results}

\subsection*{Critical temperature}

At critical temperature $T_\perp=T_{\perp c}$ one has $N_0=0$. Resolving the thermal wavelength for $T_{\perp c}$, the critical temperature is defined by the right-hand side of the last expression through the constant density $N$ and temperature anisotropy $\tau_c=T_\|/T_{\perp c}$ at $T_{\perp c}$:
\begin{equation}
T_{\perp c}= T_c\bigg[\frac{\sqrt{1/\tau-1}}{\mathrm{sinh^{-1}}\sqrt{1/\tau-1}}\bigg]^\frac{2}{3},\qquad T_c=\frac{2\pi\hbar^2}{m}\bigg[\frac{N}{g\zeta\big({\textstyle\frac{3}{2}}\big)}\bigg]^{\frac{2}{3}}
\end{equation}
where $T_c$ is the critical temperature of the isotropic boson gas with $\tau=1$. Remembering that the effect of anisotropy in the Bose distribution disappeared in the  perpendicular direction $x=0$, the temperature $T_c$ corresponds to the invariant critical temperature in the perpendicular direction. 

The expression in the brackets for $\tau<1$ can be manipulated. First the denominator is expressed as a logarithm through $\mathrm{sinh^{-1}}\sqrt{1/\tau-1}=\log\{\tau^{-1/2}[1+\sqrt{1-\tau}]\}\approx\log\{2\tau^{-1/2}(1-\tau/4)\}$, which can further be simplified. Similarly, the numerator is rewritten into some more convenient form. This yields after some simple algebra
\begin{equation}\label{eq-5}
T_{\perp c}\approx \frac{T_c}{\tau^\frac{1}{3}}\frac{\mathrm{e}^{-\frac{\tau}{3}}}{(\log 2)^\frac{2}{3}}\Big[1+\frac{\log|\tau|}{2\log2}\Big]^{-\frac{2}{3}}\lesssim1.3\frac{T_c}{\tau^\frac{1}{3}}
\end{equation}
For $\tau<1$ the exponential in the numerator is of order $O(1)$ while the additional factor is always smaller than one. The main dependence on the anisotropy is  contained in the third root of $\tau$ in the denominator.  This shows that for small $\tau$, i.e. large temperature anisotropies the effective critical temperature $T_{\perp c}$ becomes reduced like the third power of the anisotropy.

Retaining only the dominant dependence on $\tau$ and neglecting the small additional reduction by the exponential as well as the additional weakly variable  neglected terms in the denominator, the critical perpendicular temperature can be explicitly written as function of the parallel temperature 
\begin{equation}\label{eq-iso}
T_{\perp c}(T_\|)\lesssim \bigg(\frac{2\pi\hbar^2}{m}\bigg)^\frac{3}{2}\bigg[\frac{N}{g\zeta\big({\textstyle\frac{3}{2}}\big)}\bigg]T_\|^{-\frac{1}{2}}\equiv\bigg(\frac{T_c^{\:3}}{T_\|}\bigg)^\frac{1}{2}
\end{equation}
from where the effect of the anisotropy on the critical temperature is obvious. The shift $\Delta T_c=T_{\perp c}-T_c$ in the critical temperature from the isotropic state becomes
\begin{equation}
\frac{\Delta T_c}{T_c}\approx \bigg(\frac{T_c}{T_\|}\bigg)^\frac{1}{2} -1, \qquad T_\|\leq T_\perp 
\end{equation}
This is always positive and increases with decreasing parallel temperature. In an anisotropic boson gas the critical temperature for condensate formation thus shifts towards higher temperatures for any $\tau<1$. Though this shift may not be large it may become of interest in superfluidity, BEC and possibly in high-temperature superconductivity.  

\subsection*{Condensate density}

At temperatures $T_\perp<T_{\perp c}$ the chemical potential remains negative and about zero. The density of particles above ground state is thus given by Eq. (\ref{eq-dens0}), and one obtains an estimate for the density fraction in the ground-state $\epsilon_\mathbf{p}(0)$ as function of perpendicular temperature 
\begin{equation}
\frac{N_0}{N}\gtrsim\bigg(1-\frac{T_\perp}{T_{\perp c}}\bigg), \quad T_\perp <T_{\perp c} 
\end{equation}
The explicit dependence on the parallel temperature has dropped out of this expression but is implicit to $T_{\perp c}$. Similar to the isotropic case, the fraction of density in the condensate ground state depends only on the ratio of (perpendicular) temperature to critical perpendicular temperature. However, the power of the ratio is different. With decreasing perpendicular temperature the fraction of density in the condensate increases faster. Comparing with the isotropic case and using Eq. (\ref{eq-iso}) this can be written 
\begin{equation}\label{eq-conddens}
\frac{N_0}{N}\approx1-\bigg(\frac{T_\perp}{T_c}\bigg)^\frac{3}{2}\tau^\frac{1}{2}=1-\frac{N_c}{N}
\end{equation}
with $N_c$ the critical density. Now the temperature ratio has same power as in the isotropic case, it is, however, multiplied with the anisotropy factor $\tau$ which takes care of its reduction. Clearly, increasing the critical temperature allows more particles in the energy distribution to enter the condensate.  Superfluid condensates are at the heart of the microscopic theory of superconductivity. The inclusion of temperature anisotropies helps in the race for higher temperature superconductors.   
\begin{figure}[t!]\label{fig1}
\centerline{\includegraphics[width=0.5\textwidth]{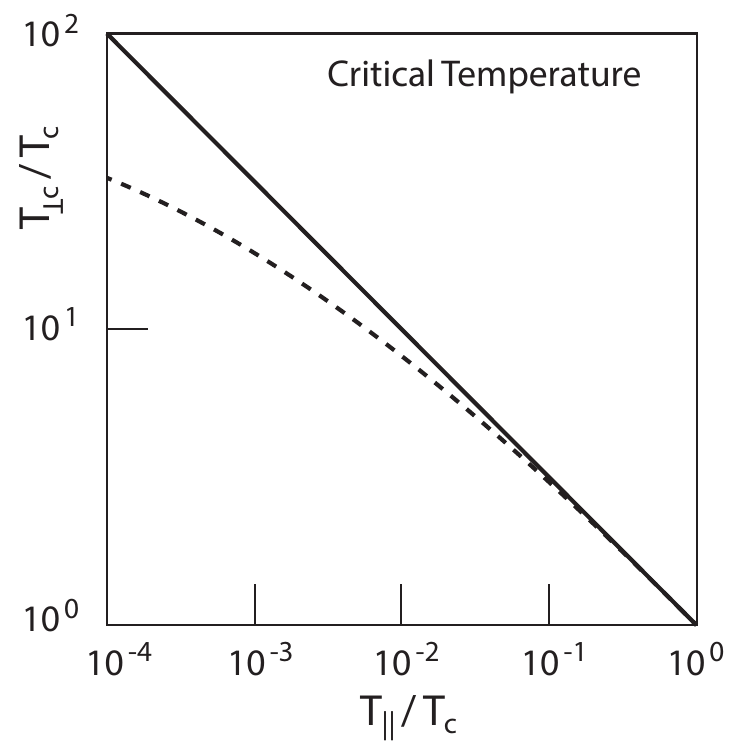} 
}
\caption[]
{Normalised critical temperature $T_{\perp c}$ as function of the normalised parallel temperature $T_\|$. Increasing  $A_\tau$ by reducing the parallel temperature below $T_\|\sim T_c$ theoretically causes a susceptible increases in the effective critical temperature $T_{\perp c}$ over its isotropic value $T_c$. The dashed line accounts for the gradual reduction of the increase that is caused by the logarithmic factor in Eq. (\ref{eq-5}). \label{crit-temp}}
\end{figure}

From a thermodynamic point of view it will be interesting to develop the full thermodynamics of an anisotropic boson gas at finite chemical potential and to also make the transition to non-ideal gases. The related questions we delegate to a separate investigation which requires returning to the grand thermodynamic potential in its integral version taking into account that the temperature is anisotropic and therefore the angular integration is implicit (but see the discussion in the Methods section where, for the ideal gas, the partition function, equation of state and number densities are given). Like in the isotropic bosonic case one can introduce new functions $g_\frac{3}{2}(z_0,\tau)$ and $g_\frac{5}{2}(z_0,\tau)$ which, for $\tau=1$, reproduce their isotropic versions. On the other hand, for $\mu=0$, they allow writing
\begin{equation}
\frac{N}{N_c}=\frac{g_\frac{3}{2}(z_0,\tau)}{g_\frac{3}{2}(1,0)}, \quad\mathrm{with}\quad {\lambda_\perp^3}N_c={g}g_\frac{3}{2}(1,0)
\end{equation}
for the ratio of density to ground state density as function of chemical potential and temperature anisotropy, which also implies that $N_0$, the ground state density (\ref{eq-conddens}), is finite for $N\lambda_\perp^3/g\leq g_\frac{3}{2}(1,0)$. This formula is the straight generalisation of the equivalent forms in an isotropic bosonic gas to finite temperature anisotropies. 

Introducing such a temperature anisotropy we thus find that the critical temperature $T_c$ can indeed be increased by some factor which depends on the anisotropy $\tau<1$ and also weakly on the chemical potential. Such an increase also implies an enhancement in the condensate number density $N_0$ of the massive bosons. Both these effects are in favour of BEC and, if survive, also of superconductivity at higher temperatures. A graphical representation of this effect on the critical temperature is given in Fig. \ref{crit-temp}. 

\section*{Discussion}

{Figure 1 shows that even a weak temperature anisotropy $\tau\lesssim1$ can in principle increase the critical temperature in a massive boson gas from a few up to a few tens thus shifting the possibility for observing BEC under maintained anisotropic conditions to higher temperatures. This might be interesting by itself in applications to  BEC.} 

{The other application we have in mind is to HTSC. Currently maximum isotropic transition temperatures $T_c$ in HTSC were reached in hydrogen-sulfides (H$_2$S), which in high-pressure conditions assume metallic properties, becoming superconductors below $T_c\sim 203$ K \citep{drozdov2015,goncharov2017}. Imposing moderate temperature anisotropies $0.1<\tau\lesssim1$ could (possibly) raise their effective critical temperature $T_{c\perp}$, in fortunate cases close to room temperatures. This is, however, speculative because it is not known yet whether H$_2$S are BCS superconductors. In bismuthates, now known to be of BCS type, \citep{wen2018} the isotropic transition temperature currently is $T_c\sim 32$ K. Raising it close to 0$^\circ$ C requires a finite though moderate temperature anisotropy $\tau\approx 0.12$ only, as suggested here. Theoretically it thus seems promising, at least for BCS superconductors, to envisage producing temperature anisotropies in order to increase $T_c$ by some factor. Whether this could become technically feasible, remains an unresolved and difficult to answer question. Our theory just shows that, given $\tau<1$, the critical temperature in the bosonic quasiparticle gas will grow. In the first place this requires technical realisation of a temperature anisotropy. In application to HTSC it, in addition, implies that at those high critical temperatures quasiparticles can still form and remain stable for sufficiently long time such that they can undergo BEC. Up to the currently achieved isotropic critical temperatures in bismuthates this seems to be the case but is not certain at higher transition temperatures under the assumed and required anisotropic conditions. }

{In spite of these  positive prospects of the present approach we express a number of reservations. {As noted in the introduction, the crucial and most problematic concerns the experimental or technical realisation and maintenance of a temperature anisotropy in an electrically neutral boson gas which is not in thermal equilibrium. }Maintaining the temperature anisotropy clearly requires an external agent. Of course, in closed many-body systems thermalisation would quickly deplete the anisotropy by causing instability and diffusion \citep{pethick2008}. To avoid thermalisation, one  turns to sufficiently dilute bose gases in thermal non-equilibrium  \citep{orioli2019}. } 

{Charged fermions as the superfluid quasiparticle pairs in BCS theory maintain the fermion properties in any temperature anisotropy. They may be vulnerable to the presence of magnetic fields exhibiting different mobilities perpendicular (Landau levels) and parallel (bounce levels) to the field. Interacting with phonons to form quasiparticles the temperature anisotropy should be conserved.  Such processes require separate investigations. We should note, however, that it is known that thermally isotropic iron-based HTSC may become vulnerable to the presence of magnetic fields in different ways by evolving  angular spatial anisotropies when in strong fields making the transition from type I to type II superconductors \citep{kordyuk2018,kushnirenko2017,kalenyuk2017,kuzmicheva2018}.} 

{Problems arise also with critical fluctuations near transition temperature. These have been studied in isotropic bosonic gases \citep{baym2000,baym2001,arnold2001,kashurnikov2001,orioli2019} causing downward shifts of $T_c$ accompanied by reductions of the condensate density $N_0$ \citep{pitaevskii2016}. Moreover, interactions in the quasiparticle gas cause additional downward shifts in $T_c$ \citep{baym1999,mueller2001,kashurnikov2001,kordyuk2018}. A more sophisticated theory of anisotropic bosonic quasiparticle gases should include both, fluctuations and interactions. The above references suggest that the shifts produced are higher order effects. They neither deplete $T_c$ substantially nor destroy the BEC. In contrast, the effect of a maintained temperature anisotropy $\tau<1$ is of first order. It may thus be expected that the anisotropic case, if it can be realised, would maintain some upward shift of the critical temperature even though becoming reduced by both critical fluctuations and interactions. A definite answer awaits a complete field theory including these effects.}  

In this letter our interest focussed on a possibility to raise the critical transition temperature in BEC and BCS theories. We took a  simplistic approach describing the superfluid of neutral bosons or fermion quasi-particles  as a simple anisotropic boson gas in thermal non-equilibrium. Both BEC and superconductivity are vastly more complicated problems. Thus our optimism might not be fully justified. However, the model itself may as well be applicable to real massive superfluid boson gases in general even if failing to be of relevance in superconductivity. 

\section*{Methods}

Since the quasi-equilibrium distribution Eq. (\ref{eq-1}) is not obvious, we provide its derivation. Assume that the dynamics of the particles in the perpendicular $\{1,2\}$ and parallel $\{3\}$ directions is different and isotropic in the perpendicular plane. {Since this generally is a non-equilibrium state, assume in addition that this difference is maintained by some external agent over sufficiently long time. In this sense the system must be open.} We  then have (non-relativistically) for the energy levels $\epsilon_{\mathbf{p}_\perp}=p^2_\perp/2m$ and $\epsilon_{\mathbf{p}_\|}=p_\|^2/2m$. Conservation of the total perpendicular and parallel energies and particle number introduces the Lagrange multipliers $\beta_\perp, \beta_\|, \lambda_{\,N}$ in  maximising the non-equilibrium entropy $S$ in the Bose ensemble
\begin{displaymath}
S=\sum_ig_i\Bigg[\frac{\beta_\perp\epsilon_{i\perp}+\beta_\|\epsilon_{i\|}-\lambda_{\,N}\mu}{\mathrm{e}^{\beta_\perp\epsilon_{i\perp}+\beta_\|\epsilon_{i\|}-\lambda_{\,N}\mu}-1}+\log\bigg(1-\mathrm{e}^{\lambda_N\mu-\beta_\perp\epsilon_{i\perp}-\beta_\|\epsilon_{i\|}}\bigg)\Bigg]
\end{displaymath}
With inverse temperatures $\beta_{\perp, \|}=1/T_{\perp,\|}$ and chemical potential $\mu$ multiplied by  $\lambda_{\,N}$, we split $\lambda_{\,N}\mu=\beta_\perp\mu_\perp+\beta_\|\mu_\|$. These forms are unchanged when writing the partition function
\begin{displaymath}
\mathcal{Z}=\prod_\mathbf{p}\Bigg[1-\mathrm{e}^{-\beta_\perp(\epsilon_{\mathbf{p}_\perp}-\mu_\perp)-\beta_\|(\epsilon_{\mathrm{p}_\|}-\mu_\|)}\Bigg]^{-1}
\end{displaymath}
Transforming $\epsilon_{\mathbf{p}_{\perp,\|}}=\epsilon_\mathbf{p}(\sin^2\theta,\cos^2\theta)$ and $\mu_{\perp,\|}=\mu(\sin^2\theta,\cos^2\theta)$, introducing $A_\tau$, and referring to $1/T_\perp$ as reference temperature and $\tau=T_\|/T_\perp<1$ the free temperature anisotropy, this expression becomes
\begin{equation}
\mathcal{Z}=\prod_\mathbf{p}\Bigg[1-\mathrm{e}^{-\beta_\perp(\epsilon_\mathbf{p}-\mu)(1+A_\tau\cos^2\theta)}\Bigg]^{-1}
\end{equation}
which gives the quasi-equilibrium equation of state 
\begin{displaymath}
P(\beta_\perp,A_\tau,\mu)V=-\beta_\perp^{-1}\sum_\mathbf{p}\log\Bigg[1-\mathrm{e}^{-\beta_\perp(\epsilon_\mathbf{p}-\mu)(1+A_\tau\cos^2\theta)}\Bigg]
\end{displaymath}
The standard definition $n_\mathbf{p}=-\beta_\perp^{-1}\partial\log\mathcal{Z}/\partial\epsilon_\mathbf{p}$ straightforwardly yields the anisotropic Bose quasi-equilibrium distribution for the average occupation number of states
\begin{equation}
n_\mathbf{p}= \Big(1+A_\tau\cos^2\theta\Big)\Bigg[\mathrm{e}^{\beta_\perp(\epsilon_{\mathbf{p}}-\mu)(1+A_\tau\cos^2\theta)}-1\Bigg]^{-1}
\end{equation}
From here, Eq. (\ref{eq-1}) follows with $\mu\to 0$ by inspection when the occupation number $n_0$ of the ground state $p=0$ is separated out. 
Replacing the sum by an integral over momentum space yields the equation of state respectively the pressure
\begin{equation}\label{eq-22}
P=\frac{T_\perp}{2\lambda_{T_\perp}^3}\int_0^1\frac{dx}{\Big(1+A_\tau x^2\Big)^\frac{3}{2}}\: g_\frac{5}{2}\Big[z(A_\tau,x)\Big]-\frac{T_\perp}{2V}\int_0^1dx\log\Big[1-z(\big|\mu\big|,A_\tau,x)\Big]
\end{equation}
where $x=\cos\theta$ and in the integrand we introduced the familiar function 
\begin{displaymath}
g_\frac{5}{2}(z) = -\frac{4}{\sqrt{\pi}}\int_0^\infty y^2dy\:\log\Big(1-z\mathrm{e}^{-y^2}\Big)= \sum_{l=1}^\infty \frac{z^l}{l^\frac{5}{2}}
\end{displaymath}
whose expansion with respect to $z$ is given by the second part in the equation. $z(A_\tau,x)=\exp\Big[-\beta_\perp|\mu|\big(1+A_\tau x^2\big)\Big]$ is a function of $x$ which is subject to integration. (Note that for bosons the chemical potential is generally negative.)
The corresponding expression for the boson density $N$ follows integrating the average occupation number $n_\mathbf{p}$ over the phase space. This yields
\begin{equation}\label{eq-24}
N-N_0=\frac{1}{2\lambda^3_{T_\perp}}\int_0^1\frac{dx}{\Big(1+A_\tau x^2\Big)^\frac{1}{2}}g_\frac{3}{2}\Big[z(A_\tau,x)\Big]
\end{equation}
with the familiar relation between $g_\frac{5}{2}$ and $g_\frac{3}{2}(z)=z\partial_zg_\frac{5}{2}(z)= \sum_{i=1}^\infty {z^l/l^\frac{3}{2}}$.

The integration in the equation of state requires solution of the integral
\begin{displaymath}
I_\frac{5}{2}(l,z_0)=\frac{z_0^l}{2l^\frac{5}{2}}\int_0^1\frac{dx\:x^{-\frac{1}{2}}\mathrm{e}^{-l\alpha A_\tau x}}{\Big(1+A_\tau x\Big)^\frac{3}{2}} =\frac{z_0^l}{2l^\frac{5}{2}}\mathrm{B}\Big(\textstyle{\frac{1}{2}},1\Big)\Phi_1\Big(\textstyle{\frac{1}{2},\frac{3}{2},\frac{3}{2}};-A_\tau,-l\alpha A_\tau\Big)
\end{displaymath}
with $z_0=\exp(-\alpha)$ and $\alpha=\beta_\perp\big|\mu\big|$. This integral is tabulated \citep{gradshteyn1965}. It can be expressed through the Beta function B$(a,b)$ and the hypergeometric series of the first kind $\Phi_1(a,b,c;d,e)$ as done in the second part of the equation. In complete analogy we obtain the integral 
\begin{displaymath}
I_\frac{3}{2}(l,z_0)=\frac{z_0^l}{2l^\frac{3}{2}}\int_0^1\frac{dx\:x^{-\frac{1}{2}}\mathrm{e}^{-l\alpha A_\tau x}}{\Big(1+A_\tau x\Big)^\frac{1}{2}} =\frac{z_0^l}{2l^\frac{3}{2}}\mathrm{B}\Big(\textstyle{\frac{1}{2}},1\Big)\Phi_1\Big(\textstyle{\frac{1}{2},\frac{1}{2},\frac{3}{2}};-A_\tau,-l\alpha A_\tau\Big)
\end{displaymath}
which appears in the expression for the density. It is this last expression which is of interest here.

With the explicit integrals we can now write down the equation of state of the anisotropic Bose gas and the density as function of $A_\tau$ respectively $\tau<1$ for arbitrary $\mu\leq0$. Before doing so we note that the last ground-state pressure term $P_0$ in Eq. (\ref{eq-22}) can indeed be done. In fact $-1<z\leq1$, and we can expand the logarithm. This yields a series of Gaussian integrals which can be integrated term by term producing
\begin{equation}
P_0=\frac{\sqrt{\pi}}{4}\frac{T_\perp}{V}\sum_{l=1}^\infty\frac{z_0^l}{\xi_l}\mathrm{erf}\big(\xi_l\big)
\end{equation}
where $\mathrm{erf}(\xi_l)$ is the error function of argument $\xi_l=\sqrt{l\alpha A_\tau}$. Collecting all terms the equation of state of the anisotropic bosonic gas reads
\begin{equation}
P-P_0=\frac{T_\perp}{2\lambda_{T_\perp}^3}\sum_{l=1}^\infty \frac{z_0^l}{l^\frac{5}{2}}\Phi_1\Big(\textstyle{\frac{1}{2},\frac{3}{2},\frac{3}{2}};-A_\tau,-l\alpha A_\tau\Big)
\end{equation}
In the same way we obtain for the density
\begin{equation}
N-N_0=\frac{1}{2\lambda^3_{T_\perp}}\sum_{l=1}^\infty \frac{z_0^l}{l^\frac{3}{2}}\Phi_1\Big(\textstyle{\frac{1}{2},\frac{1}{2},\frac{3}{2}};-A_\tau,-l\alpha A_\tau\Big)
\end{equation}
In contrast to the equation of state, it is not possible to obtain a closed analytical form for the ground state density. This is also not expected, since the density and chemical potential are reciprocally defined. Above we have given an estimate of the ground state density in the condensate. In the same spirit the ground state pressure $P_0$ is just an implicit expression which depends on the chemical potential respectively the ground state density. For $A_\tau=0$ all these expressions trivially reduce to their well-known isotropic counterparts.

In analogy to the isotropic case, the last two expressions allow for the introduction of the two anisotropic $g$-functions 
\begin{eqnarray}
g_\frac{5}{2}(z_0,\tau)&=& \frac{1}{2}\sum_{l=1}^\infty\frac{z_0^l}{l^\frac{5}{2}}\Phi_1\Big({\textstyle\frac{1}{2},\frac{3}{2},\frac{3}{2}};-A_\tau,-l\alpha A_\tau\Big)\nonumber\\
g_\frac{3}{2}(z_0,\tau)&=& \frac{1}{2}\sum_{l=1}^\infty\frac{z_0^l}{l^\frac{3}{2}}\Phi_1\Big({\textstyle\frac{1}{2},\frac{1}{2},\frac{3}{2}};-A_\tau,-l\alpha A_\tau\Big)\nonumber
\end{eqnarray}
Here, however, the relation $g_\frac{3}{2}=z_0\partial_{z_0}g_\frac{5}{2}$ becomes invalid, because the arguments of the two hypergeometric series are different.


\section*{Acknowledgements}

This paper was part of a brief visiting scientist period  at ISSI, Bern. Hospitality of the ISSI staff and interest of the ISSI directorate is acknowledged. 
\end{document}